\documentclass[aps,showpacs]{revtex4}
\usepackage{graphicx}
\def\bc{\begin{center}}
\def\ec{\end{center}}

\def\beq{\begin{equation}}
\def\eeq{\end{equation}}

\begin{document}

\title{Dynamical entanglement in coupled optical cavities
}

\author{K. Ziegler}
\affiliation{Institut f\"ur Physik, Universit\"at Augsburg, D-86135 Augsburg, Germany
}

\begin{abstract}
We study the evolution of a photonic state in a coupled pair of anharmonic cavities and
compare it with the corresponding system of two coupled harmonic cavities. The photons in
the anharmonic cavities interact with a two-level atom and are described within the
Jaynes-Cummings model. Starting from an eigenstate of the Jaynes-Cummings model with $N$
photons, the evolution of this state in the presence of photon tunneling between the 
cavities is studied. We evaluate the spectral density, the dynamics of the return as
well as the transition amplitudes and the probability for the dynamical creation of 
a N00N state. 
\end{abstract}

\maketitle






\section{Introduction}

The dynamics of isolated many-body quantum systems has been a subject of intense research in 
atomic physics during recent years, in experiment \cite{atoms} as well as in theory \cite{yukalov09}. 
This interest has two major aspects.
One is to prepare the system in a well-defined initial state and, secondly, to study its evolution
for a period of time due to quantum tunneling and particle-particle interaction. The 
preparation of the initial state as the groundstate of a certain Hamiltonian $H_0$ and the
evolution with $\exp(-iHt)$ for a different Hamiltonian $H$ involves a sudden change $H_0\to H$,
which is usually called a quench. Such a quench can be realized in atomic systems by changing
the potential wells in which the atoms are trapped \cite{trotzky08}.
For instance, bosonic atoms are prepared in a Fock state, where a definite number of atoms are 
localized in deep optical potential wells. Then the potential barrier between a pair of neighboring wells
is suddenly reduced such that the atoms can tunnel between these wells \cite{ketterle04,oberthaler05}. 
From the theoretical point of view the statistical properties of this problem have been studied intensively 
using the Hubbard model and related models 
\cite{kollath07,rigol07,manmana07,eckstein08,moeckel08,kollar08,yukalov11,yukalov11a}.
An essential element of the dynamical analysis is the evaluation of the spectral weights with
respect to the initial state \cite{roux09,rigol10}, which links directly the quantum evolution with the
spectral properties of the underlying Hamiltonian.  

Besides systems of ultracold atoms an alternative approach of controllable bosons is based on photons. 
Then the role of the potential wells is played by microwave or optical cavities. 
The experimental preparation of Fock states in a optical cavity has been achieved
recently \cite{brune08,wang08}. This is a crucial
step towards a systematic study of correlated many-body systems with photonic states. The interaction
between the photons is indirectly mediated by atoms inside the optical cavities, which interact directly with
the photons \cite{carmichael08,schuster08,ziegler12}. Once a Fock state with $N$ photons has been prepared inside 
an optical cavity, we can couple the latter with another optical cavity by a waveguide or an optical fiber.
Then the photons can tunnel between the 
two optical cavities, leading to a quantum evolution of the initial Fock state $|N,0\rangle$ within the 
Hilbert space that is spanned by the eigenstates of the Hamiltonian of the new system. This new Hamiltonian
can be approximated, for instance, by the Hubbard Hamiltonian, as suggested recently by several groups 
\cite{imamoglu97,grangier98,hartmann06,hartmann08}.
This type of system, including atomic degrees of freedom, was studied within a Hartree-Fock 
approximation \cite{ji09}. 

The evolution from the initial Fock state can, in principle, lead to an entangled state, such as the
N00N state $(|N,0\rangle +|0,N\rangle)/\sqrt{2}$. The N00N state has attracted much attention because it
can be used for highly accurate interferometry and other precision measurements 
\cite{lee02,walther04,mitchell04,afek10} and for technological application such as optical lithography 
\cite{boto00}. 
Various methods for the creation of photonic N00N states have been suggested in the literature
\cite{kok02,cable07} and indeed experimentally created up to $N=5$ photons \cite{afek10}. This
clearly indicates that the creation of these entangled states is realistic. However, it still remains
a problem to created N00N states for large $N$. The author discussed recently the dynamical creation
of a N00N state from ultracold bosonic atoms in a double well \cite{ziegler11}, which shows that 
a balanced effect of inter-well tunneling and intra-well interaction can produce such a state with
moderate probability. Since photon-photon interaction can also be mediated in a cavity by coupling
the photons to atoms, we will analyze in the following the dynamical creation of entangled (N00N) states for
a pair of anharmonic cavities, described by coupled Jaynes-Cummings models \cite{jaynes63,cummings65}.



The paper is organized as follows: In Sect. \ref{sect:def} we introduce the basic quantities
for the dynamics of our system with two cavities. Then we discuss the return probability, the
transition probability and their relation with entangled states in Sect. \ref{sect:return_etc},
applying it to a single cavity with a two-level atom (Sect. \ref{sect:single_cavity}), 
to a coupled pair of harmonic cavities (Sect. \ref{sect:2harmonic}) and eventually to a 
coupled pair of anharmonic cavities (Sect. \ref{sect:2anharmonic}). The results of the recursive
projection method of Sect. \ref{sect:2anharmonic} are discussed and compared with the results
of the coupled harmonic cavities in Sect. \ref{sect:results}. Finally, we summarize the work in
Sect. \ref{sect:discussion}.

\section{Spectral Density and the evolution of isolated systems}
\label{sect:def}

We consider a system which is isolated from the environment. In terms of photonic states this can be realized by
an ideal optical cavity. With the initial state $|\Psi_0\rangle$  
we obtain for the time evolution
$
|\Psi_t\rangle=e^{-iHt}|\Psi_0\rangle
$
or the evolution of the return probability $|\langle\Psi_0|\Psi_t\rangle|^2$ with the return amplitude
$
\langle\Psi_0|\Psi_t\rangle =\langle\Psi_0|e^{-iHt}|\Psi_0\rangle 
$.
A Laplace transformation relates the return amplitude with the resolvent through the identity
\beq
\langle\Psi_0|\Psi_t\rangle 
=\int_\Gamma \langle\Psi_0|(z-H)^{-1}|\Psi_0\rangle e^{-izt}dz
\ ,
\eeq
where the contour $\Gamma$ encloses all the eigenvalues $E_j$ ($j=0,1,...,N$) of $H$, assuming that
the underlying Hilbert space is $N+1$ dimensional.
With  the corresponding eigenstates $|E_j\rangle$ the spectral representation of the resolvent is 
a rational function of $z$:
\beq
\langle\Psi_0|(z-H)^{-1}|\Psi_0\rangle=\sum_{j=0}^N\frac{|\langle\Psi_0|E_j\rangle|^2}{z-E_j}
=\frac{P_N(z)}{Q_{N+1}(z)}, \ \ \
Q_{N+1}(z)=\prod_{j=0}^N (z-E_j)
\ ,
\label{resolvent2}
\eeq
where $P_N(z)$, $Q_{N+1}(z)$ are polynomials in $z$ of order $N$, $N+1$, respectively. 
These polynomials can be evaluated by the 
recursive projection method (RPM) \cite{ziegler10a}. The method is based on a systematic expansion of
the resolvent $\langle \Psi_0|(z-H)^{-1}|\Psi_0\rangle$, starting from the initial state $|\Psi_0\rangle$.
It can be understood as a directed random walk in Hilbert space, where each subspace
is only visited once. The latter is the main advantage of the RPM that allows us to calculate 
efficiently the resolvent 
$\langle \Psi_0|(z-H)^{-1}|\Psi_0\rangle$ on an $N+1$-dimensional Hilbert space. 

The expression in Eq. (\ref{resolvent2}) suggests the introduction of the photonic spectral density 
$\rho_{0,0}(E)$ as the imaginary part of the resolvent:
\beq
\rho_{0,0}(E)=\lim_{\epsilon\to 0}\frac{1}{\pi}Im \langle \Psi_0|(E-i\epsilon-H)^{-1}|\Psi_0\rangle
= \lim_{\epsilon\to 0}\frac{\epsilon}{\pi}\sum_{j=0}^N\frac{|\langle\Psi_0|E_j\rangle|^2}{\epsilon^2+(E-E_j)^2}
= \sum_{j=0}^N|\langle\Psi_0|E_j\rangle|^2\delta(E-E_j)
\ ,
\label{spectrald}
\eeq
where $|\Psi_0\rangle$ is a reference state. In other words, $\rho_{0,0}(E)$ is the diagonal element of the density
matrix with respect to $|\Psi_0\rangle$.
Then the return amplitude can be written as the Fourier transform of the photonic spectral density 
\beq
\langle\Psi_0|\Psi_t\rangle =\int\rho_{0,0}(E)e^{-iEt}dE
=\sum_{j=0}^N|\langle\Psi_0|E_j\rangle|^2e^{-iE_jt}
\ .
\label{return_a}
\eeq
We can also evaluate other elements of the density matrix, such as the off-diagonal element
\beq
\rho_{1,0}(E)=\frac{1}{\pi}\lim_{\epsilon\to 0}Im \langle \Psi_1|(E-i\epsilon-H)^{-1}|\Psi_0\rangle 
=\sum_j\langle \Psi_1|E_j\rangle\langle E_j|\Psi_0\rangle\delta(E-E_j) 
\ ,
\label{fourier_c2}
\eeq
whose Fourier transforms gives the transition amplitude between the states $|\Psi_0\rangle$ and $|\Psi_1\rangle$:
\beq
\langle\Psi_1|\Psi_t\rangle =\int\rho_{1,0}(E)e^{-iEt}dE
=\sum_{j=0}^N\langle\Psi_1|E_j\rangle\langle E_j|\Psi_0\rangle e^{-iE_jt}
\ .
\label{transition_a}
\eeq
To characterize the entangled state $|\Psi_t\rangle=c_0|\Psi_0\rangle + c_1|\Psi_1\rangle$ that may 
appear during the evolution, we need to evaluate the amplitudes (\ref{return_a}), 
(\ref{transition_a}) and count how often they realize certain values $c_0$, $c_1$ 
simultaneously during a long period of time. After normalization, this defines the
conditional probability $P(c_0,c_1)$ for having $\langle\Psi_0|\Psi_t\rangle=c_0$ and 
$\langle\Psi_1|\Psi_t\rangle=c_1$ at a given time $t$.



\section{Return probability, transition probability and entanglement}
\label{sect:return_etc}

The central idea is to prepare an eigenstate of the cavity as initial state and then change
the conditions of the system, either by adding a two-level atom to the cavity or by coupling
another cavity through an optical fiber. As a result of this change, the system starts to 
evolve in Hilbert space to visit all possible eigenstates of the new Hamiltonian which have 
a non-vanishing overlap with the initial state. 
During its evolution the system may visit entangled states with certain probability. The
dynamics and the entangled states will be calculated in the following subsections for 
three different cases.

\subsection{Single cavity with a two-level atom}
\label{sect:single_cavity}

An anharmonicity in a cavity can be created by adding an atom which interacts with the photons 
\cite{kleppner81,schuster08,carmichael08}.
In the case of a single two-level atom we can describe the absorption and emission
of photons by the atom approximately with the Jaynes-Cummings model \cite{jaynes63,cummings65},
whose Hamiltonian reads
\beq
H_{JC}= \omega_0 a^\dagger a +(\omega_0+\Delta) c^\dagger c -g(a^\dagger c+c^\dagger a)
\ .
\label{hamiltonian0}
\eeq
$\Delta$ is the detuning between the atomic excitation energy and the photon energy,
$c^\dagger$ ($c$) is the creation (annihilation) operator of the atomic excitation,
and $g$ is the coupling strength between the photons and the atom.
The eigenvalues of this Hamiltonian are \cite{jaynes63,cummings65}
\beq
E_{n,\pm}=\omega_0(n+1/2)\pm\sqrt{\Delta^2+4g^2(n+1)}
\label{jc_spectrum}
\eeq
with eigenstates for $\Delta=0$
\[
|n,\pm\rangle=\frac{1}{\sqrt{2}}[\pm |n:1\rangle+|n+1:0\rangle] , \ \ \ 
E_{n,\pm}=\omega_0(n+1/2)\pm 2g\sqrt{n+1}
\ ,
\]
where $|n:j\rangle$ is a Fock state with $n$ photons and an atomic state with $j=0$ (atomic groundstate)
and $j=1$ (excited atom). Thus, the eigenstates are 
superpositions of 
two Fock states, one with $n$ photons and the atom in the ground state
and one with $n-1$ photons and the atom in the excited state. This implies that the energy levels
can be doubly degenerate with respect to the number of photons. 
A superposition of many eigenstates states for the initial state $|\Psi_0\rangle$ can lead
to a more complex behavior, such as a collapse and revival dynamics \cite{puri85,buzek89}.

The eigenstates of a harmonic cavity (without the atom) are Fock states $|n\rangle$ with $n$ photons.
We prepare a harmonic cavity in one of these eigenstates.
This can be achieved in a real system, as recent experiments have demonstrated \cite{brune08,wang08}. 
Then we add a two-level atom in the ground state $|0\rangle$, such that the initial Fock state of the combined
system is a product state $|n:0\rangle\equiv|n\rangle|0\rangle$, whose evolution is described 
within the JC model as
\[
|\Psi_t\rangle=e^{-iH_{JC}t}|n:0\rangle
=e^{-iE_{n-1,+}t}\frac{1}{\sqrt{2}}|n-1,+\rangle
+e^{-iE_{n-1,-}t}\frac{1}{\sqrt{2}}|n-1,-\rangle
\ .
\]
Here we have assumed that the two-level system and the cavity are in resonance (i.e. $\Delta=0$)
in order to have an optimal exchange between the atom and the photons. This implies
for the return (transition) amplitudes 
\[
\langle n:0|\Psi_t\rangle=e^{-i\omega_0(n-1/2)t}\cos(2g\sqrt{n}t) , \ \ \
\langle n-1:0|\Psi_t\rangle=-ie^{-i\omega_0(n-1/2)t}\sin(2g\sqrt{n}t)
\]
and for the return (transition) probabilities
\[
|\langle n:0|\Psi_t\rangle|^2=\cos^2(2g\sqrt{n}t) , \ \ \
|\langle n-1:1|\Psi_t\rangle|^2=\sin^2(2g\sqrt{n}t) 
\ .
\]
Thus, we see Rabi oscillations between the two Fock states $|n:0\rangle$ and $|n-1:1\rangle$
with frequency $\Omega_R=2g\sqrt{n}$.
This is an extreme case of Hilbert-space localization, where the system is constraint to a two-dimensional
subspace. It is enforced by the fact that the eigenstates of the JC model are linear combinations
of only two Fock states. The drawback of the extreme localization in Hilbert space is that we are not
able to create dynamically entangled Fock states, except for the superposition of $|n:0\rangle$ and
$|n-1:1\rangle$.
In the general case the eigenstates of the Hamiltonian may be a superposition
of many Fock states. Then the overlap of the eigenfunctions with the initial Fock state plays a crucial role.
As a simple example we consider in the next section two harmonic cavities which are coupled by an optical fiber.

\subsection{Two coupled harmonic cavities}
\label{sect:2harmonic}

The Hamiltonian of two uncoupled harmonic cavities is
$
H_{hc}=\omega_0 \sum_{j=1,2} a^\dagger_j a_j
$,
where the index $j=1,2$ refers to the two cavities, has product Fock states 
$|N-k,k\rangle\equiv|N-k\rangle|k\rangle$ ($k=0,..,N$) as eigenstates. 
Now we couple these cavities and obtain the Hamiltonian
\[
H_{hc}=-J(a_1^\dagger a_2+a_2^\dagger a_1)+\omega_0 (a^\dagger_1 a_1+a^\dagger_2 a_2)
\]
with eigenstates  $|N-k;k\rangle$. Here we assume that $N$ is even and
calculate the overlap of the eigenstates $|N-k;k\rangle$ 
with the initial Fock state $|N,0\rangle$ \cite{ziegler12}:
\beq
\langle N,0|N-k;k\rangle=2^{-N/2}{N\choose k}^{1/2}
\ ,
\label{binomial}
\eeq
which is non-zero for all eigenstates. The density-matrix elements with respect to 
$|\Psi_0\rangle = |N,0\rangle$ and $|\Psi_1\rangle = |0,N\rangle$ are
\beq
\rho_{0,0}(E)=2^{-N}\sum_{k=0}^N {N\choose k}\delta(E+J(2k-N) , \ \ \ 
\rho_{1,0}(E)= 2^{-N}\sum_{k=0}^N{N \choose k}(-1)^k\delta(E+J(2k-N))
\ .
\label{free_bose3b}
\eeq
Thus, there is a binomial distribution for the spectral weight $|\langle N,0|N-k;k\rangle|^2$, 
with a maximal overlap for an equally distributed number of photons.
For large $N$ the binomial distribution
becomes a Gaussian distribution, where the width of the envelope is related to the energy level spacings  
$\Delta E=2J$. The Gaussian result resembles the Central Limit Theorem for 
independent photons. Such a behavior was also found previously for freely expanding bosons from an initial 
Fock state \cite{cramer08}. 
A Fourier transformation reveals a periodic behavior of the return and transition amplitudes as
\beq
\langle N,0|e^{-iHt}|N,0\rangle=\cos^N(Jt) , \ \ \
\langle 0,N|e^{-iHt}|N,0\rangle=(-i)^N\sin^N(Jt)
\ .
\label{non_int_exp}
\eeq
Thus the evolution of the Fock state is periodic with period $2\pi/J$ but leads to a 
N00N state $c_0|N,0\rangle + c_N|0,N\rangle$ only with a probability that decays 
exponentially with $N$. For larger values of $N$ the probability $P(c_0,c_N)$ indicates
an anti-correlation: $P(c_0,c_N)$ vanishes as soon as both $c_0$ and $c_N$ become nonzero.
Therefore, the overlap of $|\Psi_t\rangle$ with a N00N state is strongly suppressed. 
This is a consequence of the fact that for an increasing $N$ the 
particles disappear in the $(N+1)$--dimensional Hilbert space because there is no 
constraint due to interaction.

\subsection{Two coupled cavities with two-level atoms}
\label{sect:2anharmonic}

Now we prepare an anharmonic cavity of Sect. \ref{sect:single_cavity} in the eigenstate of the 
JC model $|N,+\rangle$ and connect it with another anharmonic cavity which is in the state 
$|0,+\rangle$. After the connection the photons start to tunnel between the two cavities. 
This system is now described by the Hamiltonian
\beq
H_{2JC}=-J( a^\dagger_1 a_2 + a^\dagger_2 a_1)
+\sum_{j=1,2}[
\omega_0 a^\dagger_j a_j +\omega_0 c^\dagger_j c_j -g(a^\dagger_j c_j+c^\dagger_j a_j)]
\ ,
\label{hamiltion2}
\eeq
where the first term describes the tunneling of photons between the cavities with rate $J$ and the second term 
represents the absorption and emission of photons by the two-level atom inside each cavity. For the 
initial state we prepare a product of JC eigenstates 
$|N,\sigma;0,\sigma'\rangle\equiv |N,\sigma\rangle |0,\sigma'\rangle$. The operators of
the Hamiltonian (\ref{hamiltion2}) act on the JC eigenstates separately. In particular, photon tunneling 
is controlled by the following matrix elements
\[
\langle k-1,+|a|k,+\rangle=\frac{\sqrt{k+1}+\sqrt{k}}{2} , \ \ \  
\langle k-1,+|a|k,-\rangle=\frac{\sqrt{k+1}-\sqrt{k}}{2} 
, 
\]
\[ 
\langle k-1,-|a|k,+\rangle=\frac{\sqrt{k+1}-\sqrt{k}}{2} , \ \ \  
\langle k-1,-|a|k,-\rangle=\frac{\sqrt{k+1}+\sqrt{k}}{2}
\]
\[
\langle k-1,+|a^\dagger|k,+\rangle=\frac{\sqrt{k+2}+\sqrt{k+1}}{2} , \ \ \  
\langle k-1,+|a^\dagger|k,-\rangle=\frac{\sqrt{k+2}-\sqrt{k+1}}{2} , 
\]
\beq
\langle k-1,-|a^\dagger|k,+\rangle=\frac{\sqrt{k+2}-\sqrt{k+1}}{2} , \ \ \  
\langle k-1,-|a^\dagger|k,-\rangle=\frac{\sqrt{k+2}+\sqrt{k+1}}{2}
\ .
\eeq
With the ratio
\beq
\frac{\sqrt{k+2}-\sqrt{k+1}}{\sqrt{k+2}+\sqrt{k+1}}\approx0
\eeq
the flipping of the atomic levels during the photon tunneling between the cavities is strongly suppressed. 
Moreover, we can use $\sqrt{k+1}+\sqrt{k}\approx2\sqrt{k}$.
With these approximations we decouple the $\pm$ states in the cavities to obtain a Hubbard-like
model, where the $n_j^2$ interaction is replaced by a $\sqrt{n_j}$ photon-photon interaction:
\beq
H_{eff}=-J( a^\dagger_1 a_2 + a^\dagger_2 a_1)+\omega_0 (a^\dagger_1 a_1+a^\dagger_2 a_2)
+2\sigma g(\sqrt{a^\dagger_1 a_1}+\sqrt{a^\dagger_2 a_2})
\ .
\label{eff_ham}
\eeq
Just like the Hubbard model, this Hamiltonian has a two-fold degeneracy for $J=0$ due to the equivalence
of the two cavities. On the other hand, the interaction is weaker than the $n_j^2$ interaction of the
Hubbard model. This indicates that the properties of the coupled JC models may resemble the behavior
of the Bose-Hubbard model in a double well \cite{ziegler11}, with less pronounced interaction features though. 

The appearance of $H_{eff}$  brings us in the position to apply the RPM of Ref. \cite{ziegler11}, 
only replacing the interaction term. Assuming that $N$ is even, all projected spaces ${\cal H}_{2j}$ 
are two-dimensional and spanned by $\{|N-j,\sigma;j,\sigma\rangle , |j,\sigma;N-j,\sigma\rangle\}$ ($j=0,...,N/2$). 
This leads to a recurrence relation in the base of the two JC states 
$|N,\sigma ;0,\sigma\rangle$, $|0,\sigma;N,\sigma\rangle$ as initial states. The value of $\sigma=\pm1$ affects
only the sign of coupling between cavity photons and the two-level system. Therefore, we ignore subsequently
the $\sigma$ dependence in the matrix elements.
If we define
\beq
a_{N/2} = \langle N,0|(z-H)^{-1}|N,0\rangle = \langle 0,N|(z-H)^{-1}|0,N\rangle
\ 
\label{fock1}
\eeq
and 
\beq
b_{N/2} =\langle 0,N|(z-H)^{-1}|N,0\rangle = \langle N,0|(z-H)^{-1}|0,N\rangle
\ ,
\label{fock2}
\eeq
$a_{N/2}$ and $b_{N/2}$ are obtained from the iteration of the recurrence relation 
(for details cf. \cite{ziegler11})
\beq
g_{k+1}=\pmatrix{
a_{k+1} & b_{k+1} \cr
b_{k+1} & a_{k+1} \cr
} , \ \ 
g_0=\frac{1}{z-{\tilde f}_0}\pmatrix{
1 & 0 \cr
0 & 1 \cr
} \ \ \ (k=0,1,...,N/2-1)
\label{2d_recursion}
\eeq
with coefficients
\beq
a_{k+1}=\frac{z-{\tilde f}_{k+1}-J^2a_k(N/2+k+1)(N/2-k)}
{\left[z-{\tilde f}_{k+1}-J^2a_k(N/2+k+1)(N/2-k)\right]^2-J^4b_k^2(N/2+k+1)^2(N/2-k)^2}
\label{coeff_a}
\eeq
\beq
b_{k+1}=\frac{J^2b_k(N/2+k+1)(N/2-k)}
{\left[z-{\tilde f}_{k+1}-J^2a_k(N/2+k+1)(N/2-k)\right]^2-J^4b_k^2(N/2+k+1)^2(N/2-k)^2}
\label{coeff_b}
\eeq
and
\beq
{\tilde f}_{k+1} 
=2\sigma g\sqrt{N/2+k+1} + 2\sigma g\sqrt{N/2-k-1}
\ .
\eeq
The recurrence relation terminates after $N/2$ steps with
\beq
g_{N/2}=
\pmatrix{
a_{N/2} & b_{N/2} \cr
b_{N/2} & a_{N/2} \cr
}
\ .
\label{termination}
\eeq

Here it should be noticed that there exists an invariance of the recurrence relation 
under the following simultaneous sign changes in Eqs. (\ref{coeff_a}) and (\ref{coeff_b}) 
\beq
z\to -z, \ \ g\to -g, \ \ a_j\to -a_j, \ \ b_j\to -b_j
\ .
\label{transf}
\eeq
This implies that a change from $\sigma=+$ to $\sigma=-$ in the initial JC states results in a mirror
image with respect to energy of $\rho_{0,0}(E,\sigma)$ and $\rho_{N,0}(E,\sigma)$:
\beq
\rho_{0,0}(E,\sigma)=\rho_{0,0}(-E,-\sigma) ,\ \ \ \rho_{N,0}(E,\sigma)=\rho_{N,0}(-E,-\sigma)
\ .
\label{image}
\eeq
Moreover, the density-matrix elements are invariant with respect to the harmonic frequency
$\omega_0$ of the cavities, except for a global energy shift. This reflects an important universality of 
the density matrix that allows us to separate the harmonic from the anharmonic properties of the cavities.

\section{results}
\label{sect:results}








The properties of two coupled anharmonic cavities in Sect. \ref{sect:2anharmonic} are characterized 
by two equivalent JC models and tunneling of photons between them. 
According to Eqs. (\ref{fock1}), (\ref{fock2}), the iteration of Eqs. (\ref{coeff_a}), (\ref{coeff_b}) 
gives us the following four matrix elements of the resolvent
\[
\langle N,0|(z-H)^{-1}|N,0\rangle ,\ \   \langle 0,N|(z-H)^{-1}|0,N\rangle , \ \ 
\langle 0,N|(z-H)^{-1}|N,0\rangle = \langle N,0|(z-H)^{-1}|0,N\rangle
\ .
\]
Moreover, according to Eq. (\ref{resolvent2}) these matrix elements are rational functions of $z$.
For $N$ photons these are lengthy expressions with $N+1$ poles. Therefore, it is convenient
to present the results as plots with respect to the energy.

\begin{figure}
\begin{center}
\includegraphics[width=8cm,height=7cm]{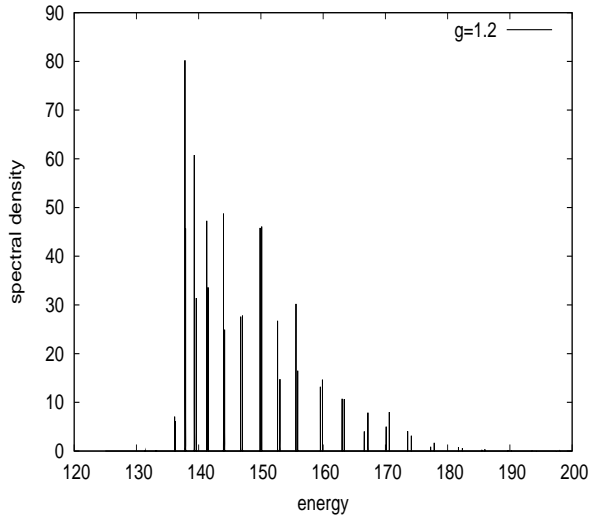}
\caption{
Spectral density $\rho_{0,0}$ 
of two coupled cavities with
two-level atoms (coupling strength $g\approx 1.2$) for 100 photons, inter-cavity tunneling rate 
$J=0.8$. All energies are measured in units of the cavity frequency $\hbar\omega_0$. 
}
\label{fig:spectrum}
\end{center}
\end{figure}

Without inter-cavity tunneling the many-photon spectrum has a two-fold
degeneracy due to the equivalence of the cavities. This degeneracy is lifted by the tunneling term, as one
can see in the spectrum presented in Fig. \ref{fig:spectrum}. Now we can compare this with the situation
of two coupled harmonic cavities, as described in Sect. \ref{sect:2harmonic} to evaluate the role of the 
photon-photon interaction. We start with the case of disconnected cavities ($J=0$) and realize that the 
spectrum for $N-k$ photons in one cavity and $k$ photons in the other cavity is completely degenerate 
for harmonic cavities
\[
E_{N-k,k}=\omega_0(N-k)+\omega_0 k=\omega_0N
\]
but only two-fold degenerate for anharmonic cavities
\[
E_{N-k,k}=\omega_0N+\sigma g(\sqrt{N-k}+\sqrt{k})
\ .
\]
After connecting the cavities the degeneracy is completely lifted and an equidistant spectrum appears with
level spacing $\Delta E=2J$ for the harmonic cavities in Eq. (\ref{free_bose3b}). $J\ne0$ also lifts
the two-fold degeneracies of the anharmonic cavities, as depicted in Fig. \ref{fig:spectrum}.
However, the levels are more irregularly distributed and their spacing is much smaller than $2J$ for pairs 
of levels. On the other hand, the spectrum does not show a spectral fragmentation, in 
contrast to the Bose-Hubbard model in a double well \cite{ziegler11}, where only in the high-energy part
of the spectrum nearly degenerate pairs of levels appear.

\begin{figure}
\begin{center}
\includegraphics[width=8cm,height=7cm]{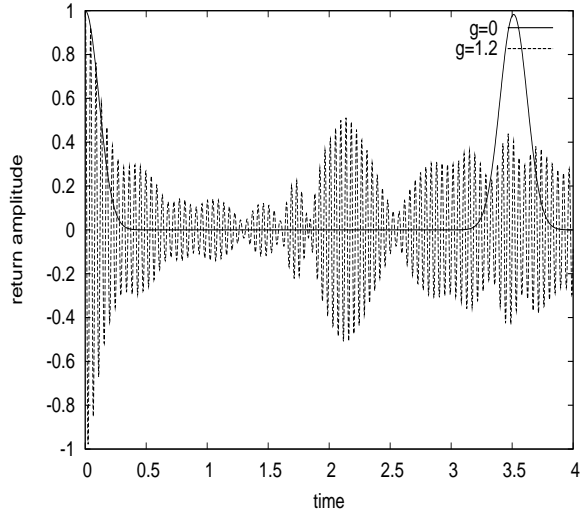}
\caption{
Return amplitude as a function of time for two coupled harmonic cavities (full curve) and for two coupled 
cavities with two-level atoms (dashed curve). 
The parameters are the same as in the previous Figure, the time is measured in units of $1/\omega_0$.
}
\label{fig:return}
\end{center}
\end{figure}

\begin{figure}
\begin{center}
\includegraphics[width=8cm,height=7cm]{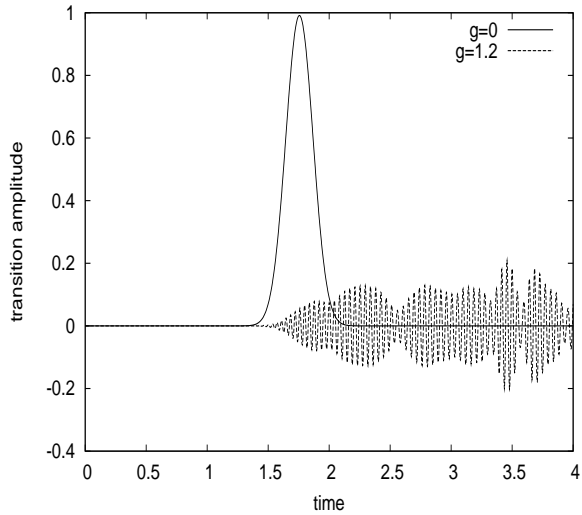}
\caption{
Transition amplitude as a function of time for two coupled harmonic cavities (full curve) and for two 
coupled cavities with two-level atoms (dashed curve). The parameters are the same as in the previous Figure.
}
\label{fig:transition}
\end{center}
\end{figure}

The difference between harmonic and anharmonic cavities is even more pronounced for the dynamics of
the return and transition amplitudes. While there is only a periodic behavior with the single frequency $J$ in
Eq. (\ref{non_int_exp}), anharmonic cavities have a more dynamic behavior (cf. Figs. \ref{fig:return},
\ref{fig:transition}). In particular, on the time scale considered in Figs. \ref{fig:return},
\ref{fig:transition}, there is no periodic behavior but oscillations on much shorter scales than 
$\pi/J\approx 4$. This is a consequence of the fact that the individual energy levels $E_k=J(2k-N)$
in Eq. (\ref{free_bose3b}) are invisible in the dynamics of the harmonic cavities due to
\beq
\sum_{k=0}^N {N\choose k}e^{iJ(2k-N)t}=(e^{iJt}+e^{-iJt})^N , \ \ \ 
\sum_{k=0}^N {N\choose k}(-1)^k e^{iJ(2k-N)t}=(-e^{iJt}+e^{-iJt})^N
\ .
\eeq
Such kind of interference effect is accidental for harmonic cavities and does not occur 
for anharmonic cavities. Therefore, we can distinguish the individual levels in the dynamics only
of the latter.

The return amplitude decays for both systems rapidly (cf. Fig. \ref{fig:return}) but it recovers much earlier
for the anharmonic cavities, not to the full value though. Remarkable is the behavior of the transition
amplitude. The time $T_t$ it takes to reach the state $|N,0\rangle$ from $|0,N\rangle$ for the first time
is about the same for both systems, indicating that $T_t\approx \pi/J$ must be solely determined by 
the tunneling rate $J$. This would allow us to measure the tunneling rate $J$ in the dynamics 
of the system, regardless of the anharmonicity. 
  
\begin{figure}
\begin{center}
\includegraphics[width=7cm,height=7cm]{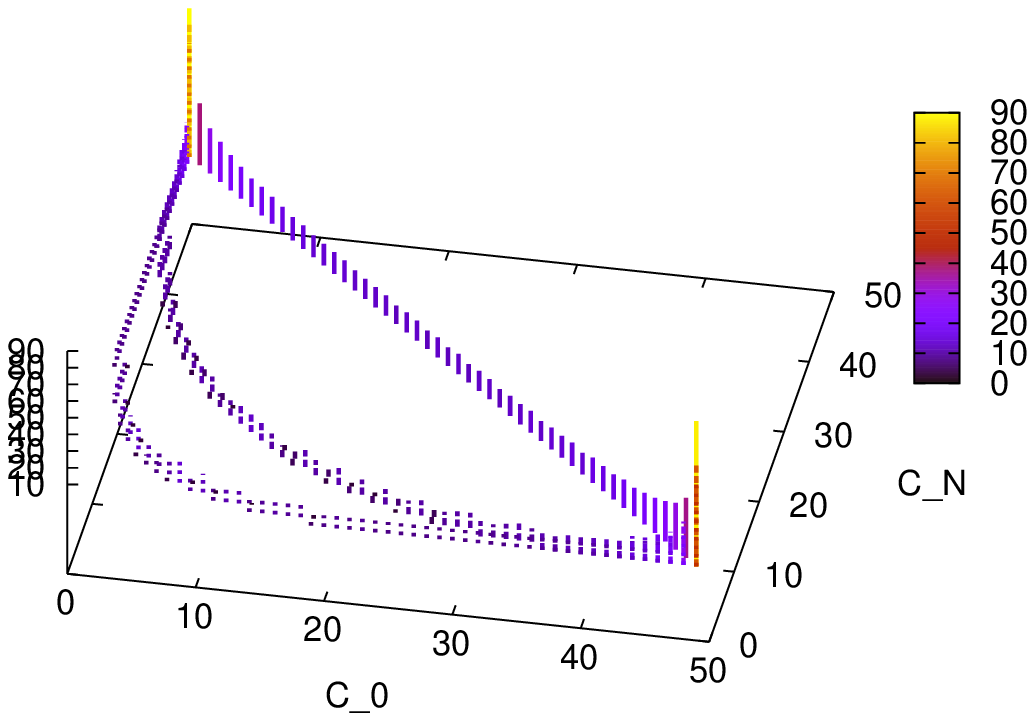}
\includegraphics[width=7cm,height=7cm]{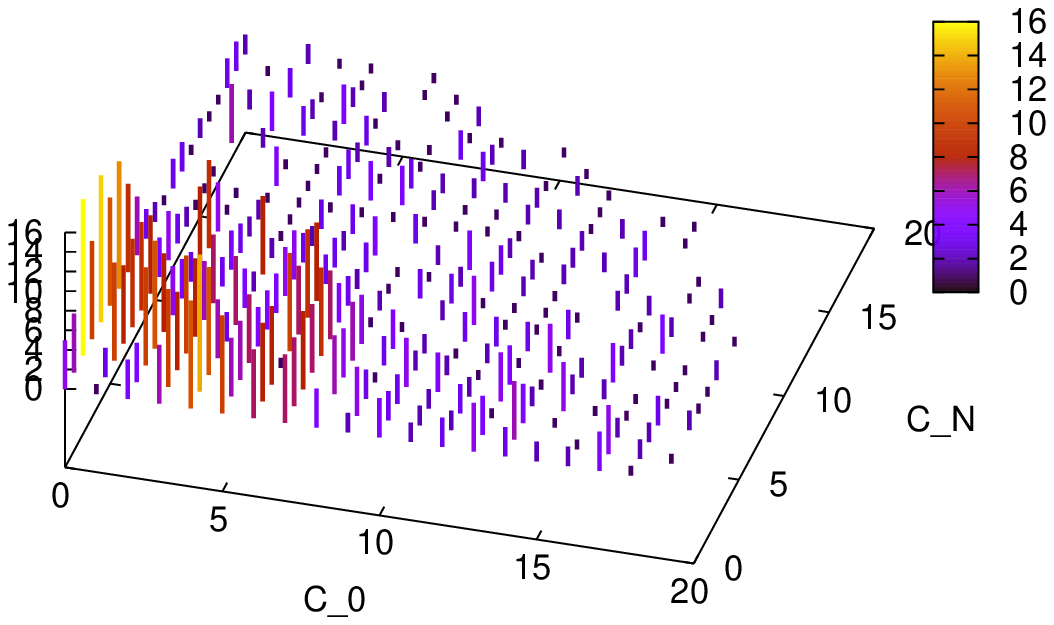}
\caption{
Probability $P(c_0,c_N)$ for creating a N00N state 
in coupled harmonic cavities with $N=2,4,6$ (from right to left in the left panel) and in coupled 
anharmonic cavities with $N=100$ (right panel). The parameters are the same as in the previous Figure.
}
\label{fig:histogram}
\end{center}
\end{figure}

Our main goal, the dynamical creation of an entangled state from a pure state, is also strongly 
affected by $T_t$, since entanglement in terms of a N00N state is not possible for times shorter than
$T_t$. For times larger than $T_t$ only the anharmonic cavities can reach the state $|N,0\rangle$ while
maintaining a non-zero overlap with the initial state (cf. Figs. \ref{fig:return}, \ref{fig:transition}).
On the other hand, only for a small number of  photons (e.g., $N=2$) the harmonic cavities are 
capable to create a N00N state dynamically.
For a discrete sequence of time steps we have counted the occurrence of certain values of the return and
transition amplitude
$(\langle N,\sigma;0,\sigma|e^{-iHt}|N,\sigma;0,\sigma\rangle , 
\langle 0,\sigma;N,\sigma|e^{-iHt}|N,\sigma;0,\sigma\rangle)$ 
for harmonic and anharmonic cavities. This yields the conditional probability 
$P(c_0,c_N)$, which was defined at the end of Sect. \ref{sect:def}. Example is plotted in 
Fig. \ref{fig:histogram}. The plots demonstrate that the dynamical creation of a N00N state is feasible 
for harmonic cavities with up to $N=4$ photons while for anharmonic cavities this can be achieved even 
for $N=100$. In comparison to bosons in a double well with Hubbard interaction \cite{ziegler11} this 
probability is quite small though. 


\section{Discussion and Conclusions}
\label{sect:discussion}

We have considered a pair of optical cavities, where the photons in each cavity couple to a two-level
atom. Then the cavities are described as JC models. For the initial state both cavities are prepared in
an eigenstate of the JC Hamiltonian. Then we have connected the cavities by an optical fiber such 
that photons can tunnel between them. The resulting evolution of the quantum state of the combined 
system is determined by an effective Hamiltonian that resembles the Bose-Hubbard model with modified
photon-photon interaction. The cavity frequency $\omega_0$ is assumed to be the same in both cavities.
In this case $\omega_0$ provides only a global shift of the spectrum, whereas the level spacing is entirely 
determined by the tunneling rate $J$ of the optical fiber. For harmonic cavities (i.e. in the absence of
the two-level atoms) the distribution of the levels with spectral weights $p_j=|\langle\Psi_0|E_j\rangle|^2$ 
is binomial with equidistant energy levels. The resulting evolution is periodic and corresponds 
to Rabi oscillations with a single frequency $J$. This behavior was observed experimentally for
weakly interacting bosonic atoms \cite{ketterle04,oberthaler05} and should also be accessible for photons 
in harmonic cavities. 
The amplitudes for visiting the initial Fock state $|N,0\rangle$ or the complimentary
Fock state $|0,N\rangle$ vary as $\cos^N (Jt)$ or $(-i)^N\sin^N (Jt)$, respectively. 
This implies for a large number $N$ of bosons that (i) these states are visited only for a very short 
period of time and (ii) the two Fock states  are visited at different times. Thus the dynamical creation
of a N00N state from a Fock state $|N,0\rangle$ is very unlikely for harmonic cavities, unless the number 
of photons is small.
The reason is that the photons can travel without seeing each other through the entire Hilbert space. 
A simultaneous overlap of $|\Psi_t\rangle$ with both Fock states $|N,0\rangle$ and $|0,N\rangle$ is 
very unlikely then. This 
is a situation in which it is very difficult to control and follow the quantum evolution. On the other hand,
applications of finite quantum systems, such as in quantum information processing \cite{garcia05,wineland11}, 
require a controllable evolution, in which only certain parts of the available Hilbert space can be visited 
with reasonable probability. In terms of our two-cavity system this means that the spectral 
weight $p_j=|\langle\Psi_0|E_j\rangle|^2$ with respect to the initial state $|\Psi_0\rangle$ is small for 
most eigenstates $|E_j\rangle$ and has only a few pronounced maxima that can be used for information storage. 
We have found that such a structured spectral density appears for anharmonic cavities, created by
coupling two-level atoms to the cavity photons. Then the photons experience a mutual influence which 
restricts their individual random walks in Hilbert space significantly and, what is even more important here, 
they can have a simultaneous overlap with both states $|0,N\rangle$ and $|0,N\rangle$.  This effect
enables the system to create dynamically a N00N state.
The latter allows us to conclude that the complex quantum dynamics of two coupled anharmonic optical 
cavities offers an approach for quantum information processing as it has also been
proposed for ultracold atoms \cite{garcia05} and cold trapped ions \cite{wineland11}.





\end{document}